\documentclass[twocolumn,showpacs,preprintnumbers,amsmath,amssymb,prb,aps]{revtex4-1}
\usepackage{epsfig}
\usepackage{epstopdf}
\usepackage{graphicx}
\usepackage{dcolumn}
\usepackage{bm}
\usepackage{slashbox}
\usepackage{textcomp}
\usepackage{subfig}
\begin{document}

\title{Fabrication of a simple apparatus for the Seebeck coefficient measurement in high temperature region}
\author{Saurabh Singh}
\altaffiliation{Electronic mail: saurabhsingh950@gmail.com}
\author{Sudhir K. Pandey}
\affiliation{School of Engineering, Indian Institute of
Technology Mandi, Kamand - 175005, India}

\date{\today}

\begin{abstract}
A simple apparatus for the measurement of Seebeck coefficient ($\alpha$) in the temperature range 300-620 K has been fabricated. Our design is appropriate for the characterization of samples with different geometries like disk and rod shaped. The sample holder assembly of the apparatus has been designed in such a way that, single heater used for sample heating purpose is enough to provide a self maintain temperature gradient across the sample. The value of $\alpha$ is obtained without explicit measurement of temperature gradient. The whole apparatus is fabricated from the materials, which are commonly available, so that any part can be replaced in case of any damage. In order to calibrate the instrument, we have carried out Seebeck coefficient measurement on nickel metal and LaCoO$_{3}$ compound. The values of $\alpha$ obtained for nickel and LaCoO$_{3}$ sample using the designed set-up are found to be equal to the values reported in the literature.

\end{abstract}


\maketitle

\section{Introduction} 
  In the past few decades, utilization of energy sources has highly been increased due to large need. The available natural resources are mostly being used to serve the purpose. The most of non-renewable natural resources (coal, petroleum and natural gases) have limited stock, and due to very fast consumption rate it may not be available in next fifty years. Therefore, the need of alternate resources of energy which can fulfill our requirement without affecting the ecological and environmental conditions are in high demand. Thermoelectric (TE) materials are one of the best resource of clean energy, which is efficient in conversion of heat energy into electrical energy. There are various applications of TE materials such as electronic component cooling, electricity production from waste heat of automobiles,  infrared sensor, etc.\cite{Bell,Snyder} The suitability of any materials for TE applications are decided on the basis of their thermoelectric efficiency. This thermoelectric efficiency is defined in terms of a dimensionless parameter called as thermoelectric figure of merit $ZT$=$\alpha$$^{2}$$\sigma$T/$\kappa$, where $\alpha$, $\sigma$, $\kappa$ and T are Seebeck coefficient, electrical conductivity, thermal conductivity and absolute temperature, respectively. The square dependent of $\alpha$ in the ZT expression makes it a very important parameter to decide the quality of materials in the TE applications. From the basic physics point of view $\alpha$ gives important information about carrier concentration $(\textit{n})$, and various phase transformation, for example of martensitic transformation in metals and alloys.\cite{Yoshida} The dependency of $\alpha$ on $\textit{n}$, and effective mass $(m^{*})$ is useful to study the electronic phase transition driven by the change in electronic band structure.\cite{Katase} Any changes around the Fermi surface due to change in $\textit{n}$ and $m^{*}$ shows the significant change in $\alpha$ versus T measurement.
  Seebeck coefficient ($\alpha$) is defined as the ratio of $\Delta$V/$\Delta$T, where $\Delta$V is the potential difference that arises due to temperature gradient $\Delta$T between hot and cold end of the sample. The accurate measurement of $\alpha$ appears to be simple in concept, but in practice it is more complex. To obtain the experimental value of $\alpha$, temperature gradient is created across the sample and induced thermoelectric voltage is measured. The accurate measurement of $\alpha$ demand the simultaneous measurement of $\Delta$T and $\Delta$V between same point of hot and cold end of the sample. There are two different way to measure the experimental value of $\alpha$, known as differential and integral method.\cite{ATBurkov,THeike} Many groups have used these two conventional methods and designed the different apparatus.\cite{Wood, Kumar, Zhou, Boffoue, Ponnambalam, Burkov, Dasgupta, Pope, Kettler, Paul, Ivory, Mishra, Rawat, Kolb} In this conventional method accurate and simultaneous measurement of $\Delta$T and $\Delta$V poses many challenges. Using this method, design of the apparatus become more complex. Ivory and Boor \textit{et al.} have proposed another method in which measurement of thermal voltages across the hot and cold ends is required and value of $\alpha$ is obtained by making the ratio of these two voltages.\cite{Ivory, Boor} In this way the error in the value of $\alpha$ due to inaccuracy of $\Delta$T measurement gets minimized, and also design of the apparatus becomes simple. This method also gives more accurate value of $\alpha$ and have certain advantages over the conventional method used by others.\cite{Wood, Kumar, Zhou, Boffoue, Ponnambalam, Burkov, Dasgupta, Pope, Kettler, Paul, Mishra, Rawat, Kolb} In spite of its simplicity, this method has been adopted by few people to fabricate the apparatus for Seebeck coefficient measurement, as this method is proposed in the recent past  years. Here, we have adopted the method given by Boor \textit{et al},\cite{Boor} and developed a simple and low cost apparatus to measure the thermopower in the temperature range 300-620 K. In our designed apparatus we have used two K-type thermocouples (TC) to measure the thermal voltages between the hot and cold end of the sample. Furthermore, the $\alpha$ value of the sample is calculated by using the formula.
\begin{equation}
\alpha = \frac{-U_{neg}}{U_{pos}-U_{neg}}\alpha_{TC}+ \alpha_{neg}
\end{equation}
where $\alpha$ is the Seebeck coefficient of sample, $\alpha$$_{TC}$ is the Seebeck coefficient of the TC, and $\alpha_{neg}$ is the Seebeck coefficient of negative leg of TC. U$_{pos}$ and U$_{neg}$ are measured voltages between positive ($\alpha$ +ve) and negative ($\alpha$ -ve) legs of the TCs, respectively. \\
The existence of small spurious voltages from the measurement system can contribute the error in Seebeck coefficient of the sample. This error comes from the presence of temperature differences in electrical connections or inhomogeneities in the thermocouples. The contribution of error due to small spurious voltage is larger when temperature gradient across the sample is less than 0.1 K. For the temperature gradient more than 0.1 K, the value of Seebeck coefficient converges to true sample's Seebeck coefficient. The correction in Seebeck coefficient value is required when the temperature gradient across the sample is less than 0.1 K and the temperature difference in the two thermocouples readings (when they are at the same temperature) is of the order of 0.001 K. If the temperature gradient across the sample is more than 0.2 K, then the contribution of the spurious voltage to the Seebeck coefficient is small. In our case, we are using the equation (6) of  Boor \textit{et al.}, for calculating the Seebeck coefficient value.  In the fabricated apparatus we can get the temperature gradient more than 1 K at the averaged sample temperature 300 K. There is self maintained temperature gradient (1-10 K) across the sample in the averaged sample temperature range 300-620 K. Therefore, at this temperature gradient value there is small spurious voltage contribution in the Seebeck coefficient of sample. The error in the Seebeck coefficient value due to small spurious voltages at large temperature gradient across the sample comes within the $\pm$ 1 micro-volt/K. This is the reason we have obtained reasonably good quality of the data for nickel metal sample and the temperature dependent behavior is similar to the data reported in the literature.\cite{Ponnambalam, Abadlia} 

  \section{Instrumentation}
  The schematic diagram of sample holder assembly of the apparatus is shown in Fig. 1a, where different components of the equipment is represented by numbers.    
\begin{figure}[htbp]
    
\begin{center}

\subfloat[]{
        \label{subfig:instrument}
        
        \includegraphics[width=0.42\textwidth]{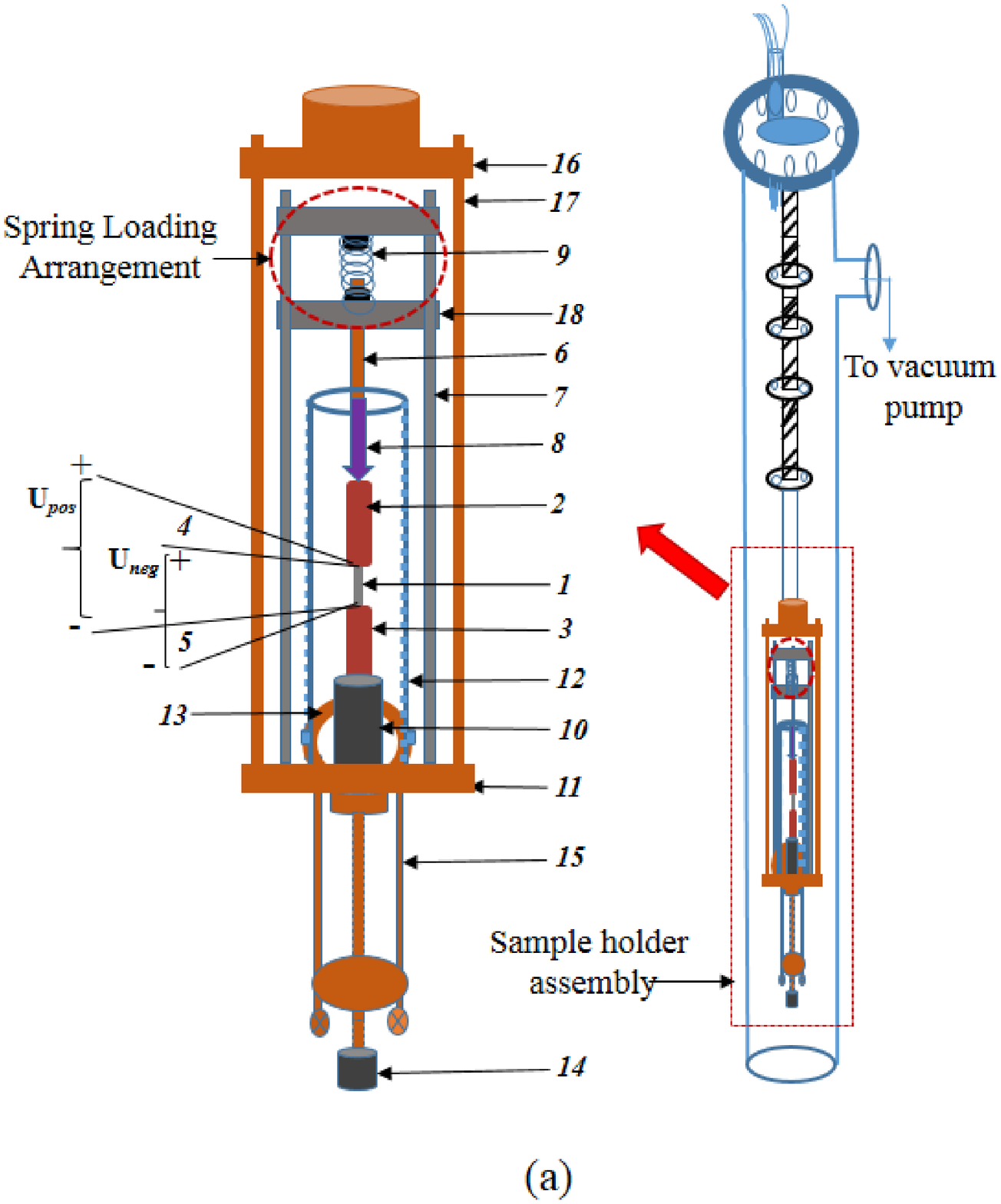} } 

\subfloat[]{
        \label{subfig:instrumentimage}
        \includegraphics[width=0.42\textwidth]{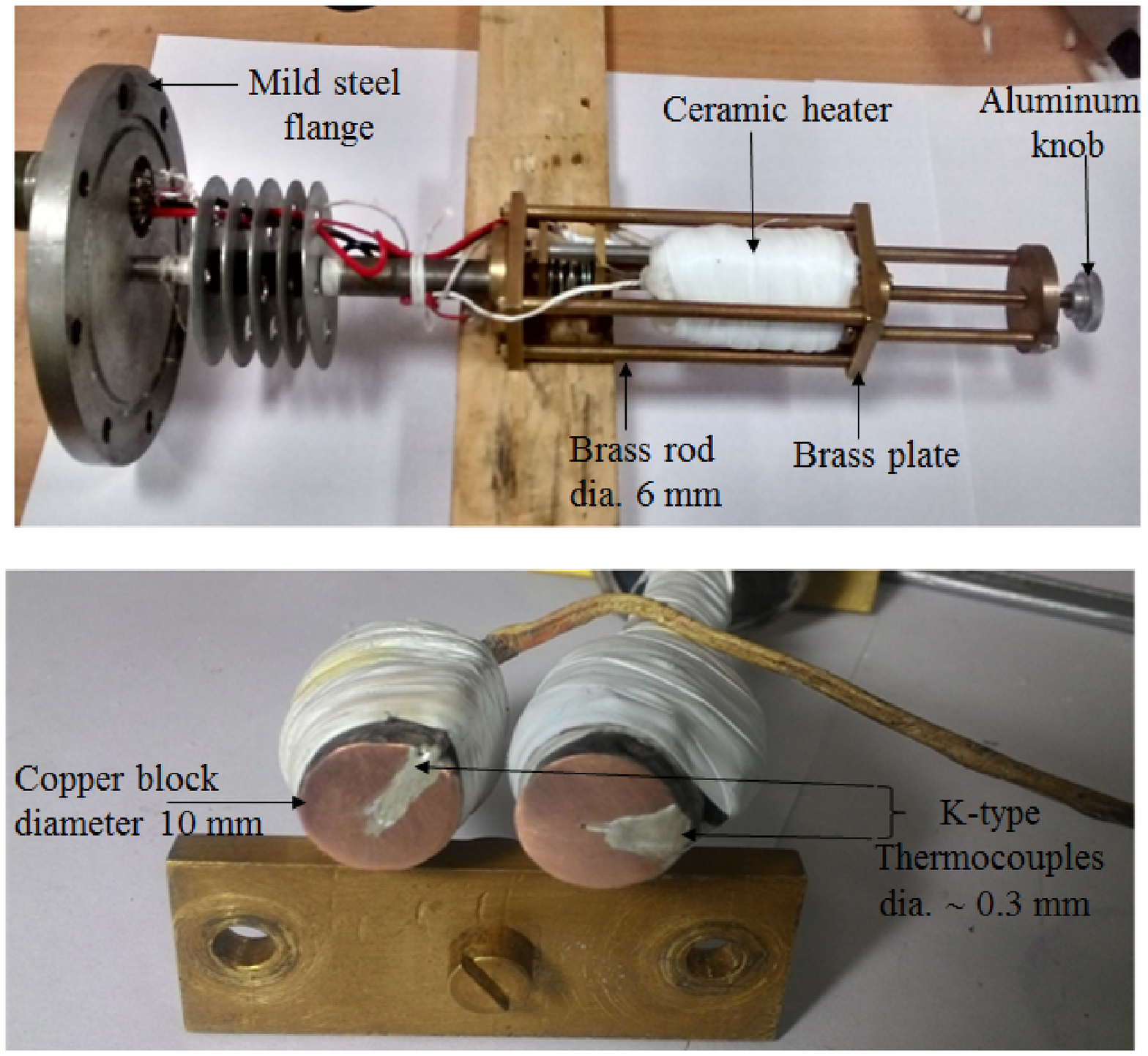} } 
\captionsetup{justification=raggedright,
singlelinecheck=false
}
\caption{(color online) Schematic diagram of the Seebeck coefficient measurement apparatus. (a) Sample holder assembly and whole apparatus arrangement, (b) Images of the measurement setup (top) and Cu blocks with region around the point of contact of the two K-type TCs (bottom).}
\label{}
  
\end{center} 
\end{figure}
A very thin ($\sim$0.3 mm diameter) K-type TCs, \textit{\textbf{\textit{4}$\&$\textit{5}}}, are attached to the Cu blocks, which are at the hot and cold end of the sample, respectively. Thermocouples are securely fixed using high temperature silver paste on the face of Cu blocks, which are very close ($\sim$0.5 mm distance) from the sample end, and it is used to measure the averaged sample temperature as well as voltages U$_{pos}$ and U$_{neg}$. The region around the point of contact of the two K-type thermocouples on each face of Cu blocks is shown in Fig. 1b. To maintain the good thermal contact throughout the measurement, we employ a spring loading mechanism, which is at top of the Cu block, \textbf{\textit{2}}, and away from the hot zone.  It is clearly shown in Fig. 1a, the back side of the copper block, \textit{\textbf{2}}, is attached with the cylindrical ceramic bead, \textit{\textbf{8}}, and this ceramic bead is fixed with the brass rod, \textit{\textbf{6}}. The Copper block, \textit{\textbf{2}}, ceramic bead, \textit{\textbf{8}}, and brass rod, \textit{\textbf{6}}, are mechanically connected back to back and finally attached with the brass plate, \textit{\textbf{18}}, using screw. This arrangement allow to maintain the required mechanical pressure on copper block, \textit{\textbf{2}}, using spring loading arrangement so that a good thermal contact is established between sample and copper blocks during the experiment.  An another copper block, \textit{\textbf{3}}, is kept on the mild steel cylindrical platform, \textit{\textbf{10}}, which is fixed mechanically in stable position on the brass plate, \textit{\textbf{11}}.\\
Sample holder assembly shown in Fig. 1a, consist of two brass plates, \textbf{\textit{11}$\&$\textit{16}}, having the dimensions of length $\sim$50 mm, breadth $\sim$40 mm, and thickness $\sim$7 mm. The bottom brass plate, \textbf{\textit{11}}, act as a base for holding the sample platform, ceramic heater and spring loading arrangement. Further, these two brass plates are well connected mechanically by using four brass rods of equal dimensions (diameter $\sim$6 mm and length $\sim$17 cm). These four brass rods act as a supporting pillar. Since these two plates are connected symmetrically to each other, therefore it provides a more stable position for sample holder assembly. The whole sample holder assembly is inserted in the vacuum chamber by using the Mild steel (MS) rod (of dimensions: length $\sim$16.5 cm, diameter $\sim$15 mm). The sample holder assembly is axially balanced by MS rod, as one end of this rod is connected at the center of the upper brass plate, \textbf{\textit{16}}, and other end of MS rod is connected to the center of vacuum flange, which is held at the top of the vacuum chamber. In this way the sample holder probe achieves more lateral stability. As shown in Fig. 1b, the whole sample probe having dimensions of $\sim$45 cm in length and $\sim$50 mm in width is inserted into the MS vacuum chamber. The cylindrical MS vacuum chamber (height $\sim$50 cm, I.D. $\sim$70 mm, O.D. $\sim$76 mm) is sufficient to prevent the sample and Cu blocks from oxidation, and it also minimizes the heat loss due to convective heat exchange. The vacuum chamber evacuated with the help of Rotary Vacuum pump, and vacuum level of 0.017 mbar is obtained. All the electrical connections are made by using the electrical feedthrough.\\
To achieve the desired temperature of the sample we have designed resistive heater, \textit{\textbf{12}}. The heater, \textit{\textbf{12}}, is made on cylindrical ceramic tube. The length of the ceramic tube is $\sim$60 mm. The I.D. and O.D. of the tube are $\sim$20 mm and $\sim$24 mm, respectively. The kanthal wire of $\sim$0.6 mm diameter is winded in $\sim$54 mm length of the tube, whereas $\sim$3 mm length at both ends are kept free, which is used for holding the heater. The Cu blocks are kept inside the ceramic tube to obtain the temperature gradient across the sample. The heater is mounted on a movable brass ring, \textit{\textbf{13}}, so that the mean position of the sample with respect to the centre of heater can be adjusted with the help of mechanical knob, \textit{\textbf{14}}.
\begin{figure}[htbp]
  \begin{center}
    \includegraphics[width=0.40\textwidth]{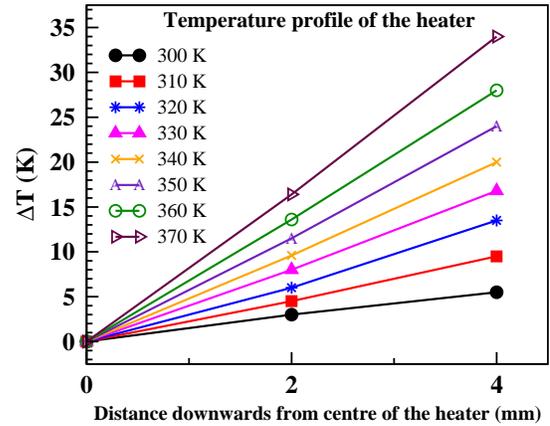}
    \label{}
    \captionsetup{justification=raggedright,
singlelinecheck=false
}
    \caption{(Color online) Temperature profile of the heater along the axial direction.}
    \vspace{0.1cm}
  \end{center}
\end{figure} 
The mechanical lifting arrangement, \textit{\textbf{15}}, helps to do initial positioning of the heater for getting the suitable temperature gradient across the sample. The temperature profile of the heater along the axial direction is shown in the Fig. 2. It is evident from the Fig. 2 that temperature gradient along the axial direction is sufficient, i.e. $>$ 1 K/mm, for accurate measurement of Seebeck coefficient of the materials. At this point it is important to note that, the heater temperature profile is measured without sample. However, in the presence of sample the values of temperature gradient across the sample are expected to be smaller. The design of the heater and sample holder probe allow to do the initial positioning of the heater such that the temperature gradient of more than 1 K across the sample at averaged sample temperature 300 K is ensure before proceeding the measurement. Also, depending on the thickness and thermal conductivity of the sample the central position of the heater with respect to the sample can be done using the mechanical knob, \textbf{\textit{14}}, attached with the heater lifting arrangement, \textit{\textbf{15}}. Although, we can do the initial positioning of the heater for suitable temperature gradient, however, the selection of appropriate thickness of the sample can also play an important role to control the temperature gradient indirectly. The temperature gradient across the sample having more thickness and low thermal conductivity will be larger.\\
The measurement procedure in our setup is quite similar to the one described by Boor \textit{et al.}\cite{Boor} To measure the $\alpha$, thermal gradient is generated across the sample. To create the thermal gradient, the temperature of the heater is raised manually by using the Crown made dual output DC Regulated Power supply (range 0-30V/5A). It is important to note that, no direct control of temperature gradient is required during the measurement process. To assure the temperature gradient across the hot and cold ends of sample, the temperature is measured using two thermocouples (\textbf{4$\&$5}). The upper Cu block, \textit{\textbf{2}}, is connected to spring loading arrangement through ceramic bead, \textit{\textbf{8}}, which acts as thermal insulator. Thus heat loss due to conduction gets minimized and heat gets shielded at hot end side of the sample. The lower Cu block, \textit{\textbf{3}}, is in physical contact with MS platform, \textit{\textbf{10}}, which is placed on the brass plate. Therefore, heat loss takes place for the cold end side of the sample through conduction process. This arrangement helps to get the self maintained temperature gradient between hot and cold end of the sample. The average sample temperature is measured manually by using the Digital Multimeter and two TCs. The U$_{pos}$ and U$_{neg}$ voltages are measured by using the same TCs connected to the two channels of Keithley 2182A nanovoltmeter.\\
The nanovoltmeter is interfaced to the computer by using the LABVIEW graphical program. The experimentally measured quantities U$_{pos}$ and U$_{neg}$ are used to calculate the value of $\alpha$ with the help of Eq. 1. The time difference between each set of the measurement for U$_{pos}$ and U$_{neg}$ is 0.2 Second. For each averaged sample temperature, we take 20 set of measurement for U$_{pos}$ and U$_{neg}$, and calculate the average value of $\alpha$. During this experimental time, the thermal gradient remains almost same and the maximum change in the average sample temperature is observed to be less than 1 K. This process minimizes the noise and improve the quality of data.\\
In order to calibrate the instrument, we have used the high purity (99.99\textdiscount) Ni metal of 6 mm diameter and about 2 mm thickness.  
      \begin{figure}[htbp]
  \begin{center}
    \includegraphics[width=0.45\textwidth]{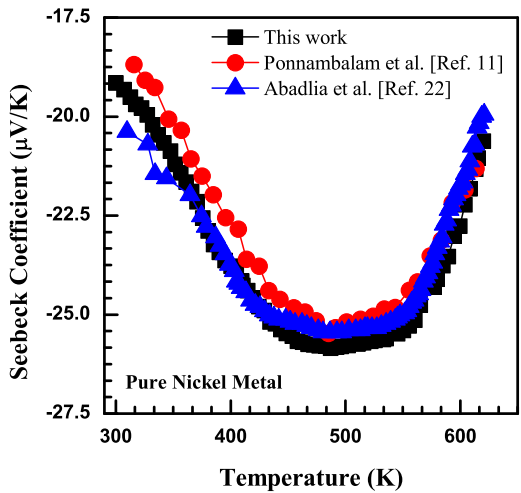}
    \label{}
    \captionsetup{justification=raggedright,
singlelinecheck=false
}
    \caption{(Color online) Seebeck coefficient $(\alpha)$ of pure nickel as a function of temperature.}
    \vspace{0.1cm}
  \end{center}
\end{figure} 
The experimental data have been collected in the temperature range 300-620 K in the interval of $\sim$5 K. The temperature dependence of $\alpha$ is shown in the Fig. 3. From the $\alpha$ vs T graph, we observe that the value of $\alpha$ at 300 K is $\sim$ -19.15 $\mu$V/K and it decreases continuously to the minimum value of $\sim$ -25.84 $\mu$V/K at $\sim$487 K. The change in the value of $\alpha$ is about 7 $\mu$V/K in 300-487 K temperature range. The value of $\alpha$ sharply increases above 487 K, and it has the value of $\sim$ -20.62 $\mu$V/K at about 620 K. A change of $\sim$ 5 $\mu$V/K is observed in the temperature range 487-620 K. To compare the temperature dependent behavior of $\alpha$ obtained from our set up with the data reported in the literature, we have plotted the literature data in the same graph. The data obtained by Ponnambalam \textit{et al.}, and Abadlia \textit{et al.}, are inserted in to the Fig. 3 by extracting the data point value using the digitization technique.\cite{Ponnambalam,Abadlia} It is important to notice that digitization process always have an inherent error in data selection from the literature graph. In our case, we expect the maximum error of about 0.5 $\mu$V/K in the literature data due to digitization. Therefore, the difference of $\sim$1 $\mu$V/K observed in between the literature data and data obtained by our setup may be due to contribution of digitization error. The temperature dependent behavior of $\alpha$ and change in its magnitude in the temperature range 300-620 K is found to be similar to those reported in the literature.\cite{Ponnambalam,Abadlia} In comparison to literature data the maximum deviation obtained in our data obtained for the nickel sample is less than $\pm$1 $\mu$V/K in the temperature range 300-620 K. Our designed setup is useful to characterize the materials in disk or rod shaped. The materials used for the measurement by our setup have the pallet having thickness from 0.5 mm to 2 mm and diameter of 5 mm to 10 mm. It is our believe that other samples of disk or rod shaped with similar dimensions can be well characterized. The present sample probe design is appropriate for sample with cross section comparable with the two surrounding copper blocks (in order to minimize the thermal contact resistance). At this point, it is important to note that the cross sectional diameter of the Cu blocks, \textbf{2} $\&$ \textbf{3}, is $\sim$10 mm and the sample (nickel) cross section is $\sim$6 mm. The sample cross section is about 4 mm less than the cross section of two Cu blocks.The factor which can lead to thermal contact resistance between sample and copper block are the roughness of the copper block and sample surface; and loose contact between the Cu block and the sample. These contact resistance affects the quality of data and one can minimize the thermal contact resistance by using the diameter of copper blocks almost equal to the diameter of the sample. Moreover, a small mismatch of radius of sample and copper block of $\sim$1-2 mm do not give a significant off-axial temperature gradient under thermal equilibrium. However, in the present case we obtained good quality data. Moreover, the two Cu blocks surrounding the sample are replaceable. Thus one can always replace these two Cu blocks with the another set of Cu blocks of suitable cross-section wherever one faces the problem of thermal contact resistance. In current design there is also possibility that the fragile sample may break. A sample of low thermal conductivity is more suitable for characterization as it have  temperature gradient (1-10 K) across the sample. However, the movable platform for heater allow us to set the sample center position to get the suitable temperature gradient. The sample with low thermal conductivity will have a large temperature gradient when the averaged sample temperature is above 100 $^{0}$C. The large temperature gradient will give the error in the averaged sample temperature as it is the average of the two temperatures i.e. hot and cold end of the sample. Thus any electronic phase transition occurs in the narrow temperature range will not be probe accurately. The measurement without any sample gives a value of 0.5-1.0 $\mu$V/K in the temperature range 300-620 K, therefore thermoelectric materials having higher value of $\alpha$ can well be characterized by using this setup.\\
  Now we are going to use the designed apparatus to study the thermoelectric property of LaCoO$_{3}$ samples with different crystallite sizes. We have chosen the LaCoO$_{3}$ sample due to its semiconducting to metal transition behavior about 540 K.\cite{RRHeikes} This compound belongs to thermoelectric oxides materials and have very large variations in thermopower in the temperature range 300-600 K.\cite{RRHeikes} The thermopower measurement of oxide thermoelectric materials will give more versatility and characterization capability of the fabricated setup. This will also help to investigate the thermoelectric materials and study of the electronic phase transition in the high temperature region.\\      
\section{Experiments}
  LaCoO$_{3}$ samples were prepared by using single step solution combustion method.\cite{Pandey} The stoichiometric amount of Lanthunum oxide (La$_{2}$O$_{3}$, from Sigma Aldrich) and Cobalt(II) nitrate hexahydrate (Co(NO$_{3}$)$_{2}$.6H$_{2}$O), from Merck) were used as starting materials. First, La$_{2}$O$_{3}$ powder was dissolved in dilute HNO$_{3}$ and then Co(NO$_{3}$)$_{2}$.6H$_{2}$O added to obtain the metal nitrates clear solution. Then, an appropriate amount of glycin (C$_{2}$H$_{5}$NO$_{2}$, from Merck) was added to the solution, which act as fuel. The added amount of glycin was such that the ratio of metal ions to that of glycin was 1:2. Further, this solution was heated at temperature around 300$^{0}$C. The combustion mixture boils, undergoes dehydration following combustion with flame, and forms the fine powders of LaCoO$_{3}$. Further, this fine powder was ground and calcined at 300$^{0}$C for 5 hour in air. The resulting powder was further ground and pressed into 5 mm circular pellets under the pressure of 25 Kg/cm$^{2}$. To obtain the samples with different crystallite size, pellets were further sintered at 800$^{0}$C, 1000$^{0}$C, 1100$^{0}$C, and 1200$^{0}$C for 24 hour in air followed by furnace cooling to the room temperature. The sintered pellets at  800$^{0}$C, 1000$^{0}$C, 1100$^{0}$C and 1200$^{0}$C are coded as LCO800, LCO1000, LCO1100, and LCO1200 respectively.\\
  The crystal structure characterization of the synthesized powder samples were determined by using the Rigaku X-Ray Diffractometer. The data were collected using CuK$_{\alpha}$ radiation at room temperature in the step scanning mode (2$\theta$ = 0.02$^0$) over the angular range of 2$\theta $ = 20$^0$-100$^0$ with the scan rate of 2$^0$/min. The high temperature measurement of the thermoelectric power was carried out between 300-600 K by using  this setup. For high temperature thermoelectric power measurements, all the samples were used in the pellets form having the same dimensions (Diameter $\sim$5 mm, thickness $\sim$0.5 mm).
 
\section{Results and Discussion}
Fig. 4 shows the  room-temperature X-ray diffraction (XRD) patterns of the LaCoO$_{3}$ powder samples, which were sintered in air at different temperatures 800$^0{C}$,1000$^0{C}$,1100$^0{C}$, and 1200$^0{C}$.
  \begin{figure}[htbp]
  \begin{center}
    \includegraphics[height=8cm, width=0.52\textwidth]{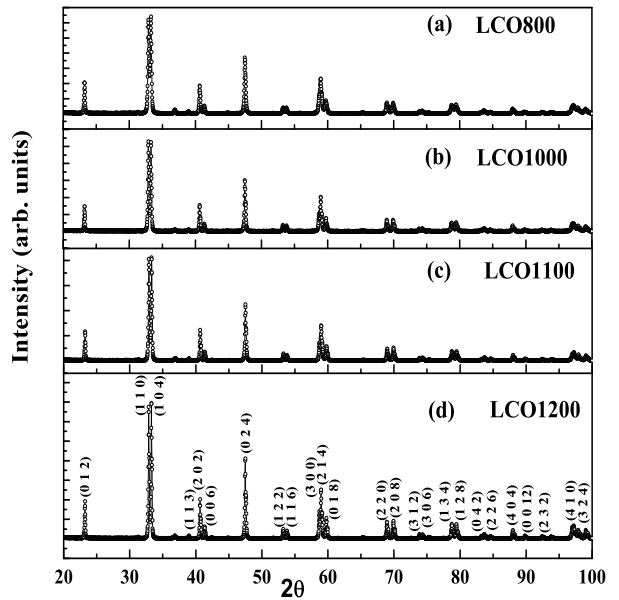}
    \label{}
    \captionsetup{justification=raggedright,
singlelinecheck=false
}
    \caption{Room temperature X-ray diffraction pattern of LaCoO$_3$ compounds. Shown are (a) LCO800, (b) LCO1000, (c) LCO1100, and (d) LCO1200.}
  \vspace{0.1cm}
  \end{center}
\end{figure}
For all the four samples i.e. LCO800, LCO1000, LCO1100, and LCO1200, we found all the peaks were present as reported in the literature.\cite{SMZhou} All the diffraction peaks are indexed corresponding to rhombohedral crystal structure described by space group R$\bar{3}$c. An unindexed peak corresponding to the Co$_{3}$O$_{4}$ phase is found at 2$\theta$ value equal to $\sim$36.82$^{0}$, whose intensity is negligibly small in comparison to the most intense peak. We have also estimated the crystallite size of the synthesized LaCoO$_{3}$ powder samples from the half-width (FWHM- Full Width Half Maximum) of peak using the Debye Scherrer formula.\cite{Warren}\\

 D = $\frac{\textit{k}\lambda}{B(2\theta).Cos(\theta)}$\\
 
 where D is the crystallite size, k is a numerical constant (0.94), $\lambda$ is the radiation wavelength of the Cu$_{K\alpha}$ X-ray source (1.5418 \AA), B(2$\theta$) is the full width in radians subtended by the half maximum intensity width of the powder pattern peak at diffraction angle 2$\theta$. To obtained the FWHM, peak is selected at 2$\theta$ = 23.20$^{0}$ for each sample and Gaussian fitting curve is used to get the best fitting. The calculated crystallite size for LCO800, LCO1000, LCO1100, and LCO1200 are about 96, 103, 110 and 121 nm, respectively. The variation of crystallite size with sintered temperature is shown in Fig. 5. From the Fig. 5, we observed that crystallite size increases with the function of sintering temperature.  The observed increment in the crystallite size is about 26 \textdiscount with change in sintering temperature from 800$^{0}$C to 1200$^{0}$C. This increment in the crystallite size with sintering temperature is due to formation of intergranular bonds and the pellets become more dense.\\
 \begin{figure}[htbp]
  \begin{center}
  \vspace{1.0cm}
    \includegraphics[width=0.30\textwidth]{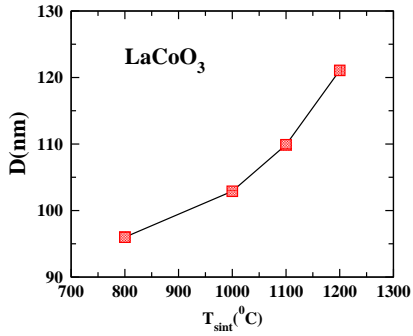}
    \label{}
    \captionsetup{justification=raggedright,
singlelinecheck=false
}
    \caption{(Color online) Crystallite size, D(nm), of LaCoO$_{3}$ compound as a function of sintering temperature T$_{sint}$ ($^{0}$C).}
    \vspace{0.1cm}
  \end{center}
\end{figure} 
  The temperature dependence of the Seebeck coefficient ($\alpha$) of the samples under investigation is shown in Fig. 6. The observed values of alpha about 300 K for the LCO1000 and LCO1100 samples are similar to those reported in the literature.\cite{SenarisRodriguez,Jirak, AMaignan} For the LCO1200 sample, we have obtained the similar temperature dependent behavior as reported by Li Fu \textit{et al.}, but a difference is observed in the magnitude of $\alpha$ value about 325 K.\cite{LiFu} In data reported by  Li Fu \textit{et al.}, the temperature interval is about 25 K, in comparison to this, we have obtained the better quality of data in the temperature interval of about 5 K.
  \begin{figure}[htbp]
  \begin{center}
    \includegraphics[width=0.40\textwidth]{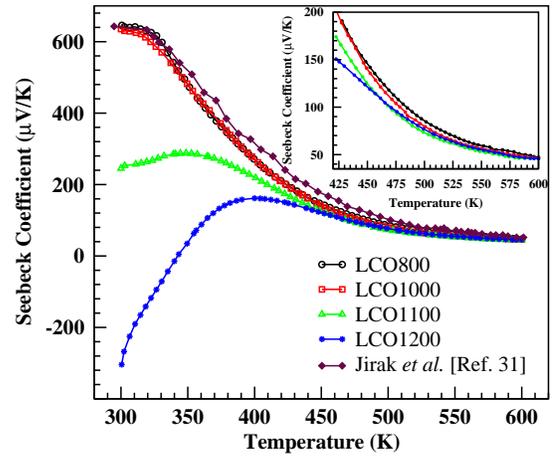}
    \label{}
    \captionsetup{justification=raggedright,
singlelinecheck=false
}
    \caption{(Color online) Temperature variations of seebeck coefficient ($\alpha$) for LaCoO$_{3}$ samples sintered at various temperatures.}
  \end{center}
\end{figure} 
For samples LCO800, LCO1000 and LCO1100, the value of $\alpha$ is found to be positive in the temperature range 300-600 K. The most striking observation is that sample sintered at 1200$^{0}$C (LCO1200) showed a change of sign of $\alpha$ in 300-345 K temperature range. The observed value of $\alpha$ at 300 K for LCO800, LCO1000, LCO1100, and LCO1200 are 645, 635, 245 and -304 $\mu$V/K, respectively. The temperature dependent behavior as well as $\alpha$ value for LCO800 and LCO1000 sample is almost same in the 300-600 K temperature range. For comparison purpose, we have plotted the literature data obtained by the Jirak \textit{et al.}\cite{Jirak} The temperature dependent behavior is similar to the reported data. There is difference between our and literature data for LCO800 and LCO1000 samples. This difference in the magnitude of $\alpha$ may be due to inherent error in data digitization, as mentioned earlier in the case of nickel. In comparison to Jirak \textit{et al.} data, the difference in the magnitude of $\alpha$ and change in temperature dependent behavior may be due to the sample synthesis conditions or change in oxygen stoichiometric ratio. The thermopower value is very sensitive to the carrier concentration, therefore synthesis conditions can highly affect the magnitude of $\alpha$. From $\alpha$ vs T graph it is clearly observed that for LCO800 and LCO1000 the value of $\alpha$ decreases continuously as the sample temperature increases. Early work of Heikes \textit{et al}, on the electric transport properties of LaCoO$_{3}$ compound shows that this materials have semiconductor to metal transition about 540 K.\cite{RRHeikes} The value of $\alpha$ is higher in the semiconducting phase, and a large variation is observed in the temperature range 300-540 K. This variation is expected in the semiconducting phase due to change in the number of carrier concentrations with temperature. In the metallic phase i.e. above 540 K, there is a small variation ($\sim$2 $\mu$V/K) in the value of $\alpha$, and almost constant upto 600 K. To show the semiconductor to metal transition more clearly, an inset figure in the Fig. 6 has been included. For LCO1100 sample, the value of $\alpha$ increases from 245 $\mu$V/K at 300 K to 287 $\mu$V/K at 350 K and further it decreases continuously. LCO1200 sample exhibits the negative value of $\alpha$ at 300 K and as temperature increases it becomes positive at about 345 K, with further increase in temperature it reach to maximum value of 160 $\mu$V/K at 400 K and then start decreasing at higher temperature. The experimentally observed values of $\alpha$ and its sign show that the nature of charge carrier in the LCO800, LCO1000, and LCO1100 samples is hole-like in the entire temperature range under investigation. However, the type of majority charge carriers for LCO1200 sample are electron-like and hole-like in the temperature range 300-345 K and 345-600 K, respectively. \\
  Although, the change in crstallite size is about 7 nm from LCO800 to LCO1000, but temperature dependent behavior of $\alpha$ is similar for both the samples. This shows that the change in the crystallite size does not have any significant effect on thermoelectric power of material. To see the effect of density of the materials on thermoelectric power, Lu Fi \textit{et al.}, have studied the LaCoO$_{3}$ compound with different densities.\cite{LiFu} They have observed that the value of $\alpha$ is almost same for the samples with two different densities. This shows that $\alpha$ has not much effect on the thermoelectric power. In comparison to LCO1000 sample, the value of $\alpha$ at 300 K for LCO1100 sample is about 390 $\mu$V/K smaller and have the same sign, whereas for LCO1200 sample it is about 939 $\mu$V/K smaller and have the negative sign. This large decrements in the value of $\alpha$ for LCO1100 and LCO1200 and change in the sign for LCO1200 may suggest that these two samples are non stoichiometric. The samples sintered at and above 1100 $^{0}$C have been reported to be oxygen deficient.\cite{Minh} Thus, also we expect that LCO1100 and LCO1200 should be oxygen deficient. The oxygen deficient compounds have excess of electrons and these electrons will also contribute to the total observed value of $\alpha$, gives rise overall decrease in the value of $\alpha$. The observation of systematic decrease in the $\alpha$ value with sintering temperature suggest that  electrons are introduced in the oxygen deficient samples. This decreasing tendency of $\alpha$ with decreasing oxygen content is also observed by Ohtani \textit{et al.}\cite{Ohtani} The LCO1200 sample is expected to be more oxygen deficient in comparison to LCO1100 and this has more free electrons to contribute to $\alpha$ value. About 300 K, the excess electrons in LCO1200 sample are majority carriers and gives the negative value of $\alpha$. The influence of oxygen deficiency on the thermoelectric behaviour is more effective in the semiconducting phase. However, above 540 K the value of $\alpha$ for all the sample sintered at different temperatures shows positive value and have not much differences. \\
    For collecting the data we have used the digital multimeter and nanovoltemeter as it is available in our laboratory. A simple power source is used to supply the current to the heater. We have used the low cost and commonly used digital multimeter for the hot and cold end temperature measurement, which is an alternative of the high cost temperature controller, as we need to measure only the averaged sample temperature. At present, the estimated overall cost of Seebeck coefficient measurement apparatus is approximately 4500 \$ ($\sim$3,00,000 INR). The major contribution in overall cost of the apparatus comes from the nanovoltmeter (approximately 3400 \$). This nanovoltmeter can be replaced by micro-voltmeter (which has an approximate cost of 900 \$ i.e. nearly four times less than the cost of nanovoltmeter), as in most of the thermoelectric materials, the thermal voltages across the sample are generally in the micro-volt range.  Thus, one can use micro-voltmeter instead of nanovoltmeter to reduce the over all cost of the measurement setup by half. In this way, the apparatus will be more cheaper.
\section{Conclusions} 
In conclusion, we have designed and fabricated a low cost apparatus for Seebeck coefficient measurement in the high temperature range. The design of sample holder assembly is very simple and it provides the self maintained temperature gradient across the sample by using a single heater. The fabricated instrument is calibrated by using the standard nickel metal. The measurement on the standard Ni sample in the temperature range 300-620 K gives the similar temperature dependent behavior as reported in the literature.\\ 
To verify the characterization capability of the designed instrument we have studied the thermoelectric behavior of LaCoO$_{3}$ compound using the fabricated setup. LaCoO$_{3}$ sample was prepared by solution combustion method. The crystalline size obtained by using Debye Scherrer formula is found to increase with sintering temperature. The temperature dependent thermopower behaviour have been studied for samples sintered at 800,1000,1100, and 1200$^{0}$C. The thermopower behaviour of all the samples were studied in the temperature range 300-600K. The $\alpha$ value at 300 K is observed $\sim$645, $\sim$635, $\sim$245 and $\sim$-304 $\mu$V/K for the samples sintered at 800$^{0}$C, 1000$^{0}$C, 1100$^{0}$C, and 1200$^{0}$C, respectively. However, at higher temperature, i.e. about 600 K, $\alpha$ value is almost same for all the samples. At 600 K, value of $\alpha$ is observed $\sim$46 $\mu$V/K for all the samples. The high temperature thermoelectric behaviour shows that the value of $\alpha$ and its sign changes with sintering temperature. This observation of large positive and a large negative thermoelectric power in undoped LaCoO$_{3}$ samples may be due to oxygen off-stoichiometry as the samples were prepared in air atmosphere. In this work it has been observed that centred of oxygen stoichiometry can open a new opportunity for tuning the thermoelectric behavior of the oxide compounds, which will be more useful for high temperature thermoelectric applications.

\end{document}